\documentclass[
 preprint,
superscriptaddress,
 amsmath,amssymb,
 aps,
prb,
floatfix
]{revtex4-2}

\usepackage{graphicx}
\usepackage{dcolumn}
\usepackage{bm}
\usepackage{color}
\usepackage{textcomp}
\usepackage{hyperref}
\hypersetup{backref=true,pdfnewwindow=true,linkcolor=black,colorlinks=true,anchorcolor=black,citecolor=black,filecolor=black,menucolor=black,urlcolor=black}

\begin{document}

\title{Growth of aligned and twisted hexagonal boron nitride on Ir(110)}

\author{Thomas Michely}
\author{Jason Bergelt}
\author{Affan Safeer}
\author{Alexander B\"ader}
\author{Tobias Hartl}
\author{Jeison Fischer}
\email{jfischer@ph2.uni-koeln.de}
\affiliation{II. Physikalisches Institut, Universit\"at zu K\"oln, Z\"ulpicher Stra\ss{}e 77, 50937 K\"oln, Germany}

\date{\today}

\keywords{}

\begin{abstract}
The growth of monolayer hexagonal boron nitride (h-BN) on Ir(110) through low-pressure chemical vapor deposition is investigated using low energy electron diffraction and scanning tunneling microscopy. We find that the growth of aligned single hexagonal boron nitride on Ir(110) requires a growth temperature of 1500\,K, whereas lower growth temperatures result in coexistence of aligned h-BN with twisted h-BN The presence of the h-BN overlayer suppresses the formation of the nano-faceted ridge pattern known from clean Ir(110). Instead, we observe the formation of a $(1 \times n)$ reconstruction, with n such that the missing rows are in registry with the h-BN/Ir(110) moir\'{e} pattern. Our moir\'{e} analysis showcases a precise methodology for determining both the moir\'{e} periodicity and the h-BN lattice parameter on an fcc(110) surface.
\end{abstract}

\maketitle

\section{Introduction}

Hexagonal boron nitride (h-BN) has made a steep career in the recent two decades. As a few-layer 2D material, it has become an indispensable insulating element in nearly all 2D heterostructures for fundamental research and application \cite{Dean10,Zhang17,Moon22}. Additionally, h-BN exhibits exciting properties as a robust room temperature single photon emitter \cite{Tran16,Moon22} and as a catalyst \cite{Wu20,Han20}. 

Following the seminal work by Nagashima et al. \cite{Nagashima95,Nagashima95b} also monolayers of h-BN grown on hot single crystal metal surfaces by decomposition of molecular precursors (chemical vapor deposition, CVD) has sparked both fundamental and applied interest. A recent review by Auwärter \cite{Auwaerter19} provides an overview of the growth and use of h-BN monolayers on single crystal surfaces as templates for atoms, molecules, and clusters as well as its use in supramolecular chemistry.

Monolayer h-BN growth on symmetry matching and densely packed fcc(111) and hcp(0001) surfaces has been in the focus of the research and thoroughly investigated. With decreasing interaction strength between the layer and the metal substrate, a transition from orientational epitaxy (aligned densely-packed substrate and zigzag direction of h-BN) to random orientation is found \cite{Auwaerter19}. However, less attention has been given to the CVD growth of monolayer h-BN on non-symmetry matching substrates of fourfold and twofold symmetry. 

In the case of twofold symmetric fcc(110) surfaces, previous studies have examined the growth on (110) surfaces of Ni \cite{Greber06}, Cu \cite{Herrmann18}, Rh \cite{Galera18,Galera19b}, Pd \cite{Corso05}, and Pt \cite{Achilli18,Steiner19,Thaler20}. Among these systems, only h-BN/Pt(110) exhibits a unique epitaxial relation with the zigzag (densely-packed) direction aligned parallel to the densely-packed $\left[1 \bar{1} 0 \right]$ direction of the substrate (in the following: aligned h-BN). Ni was interpreted to display several well-defined phases \cite{Greber06}, while the other three systems show h-BN zigzag orientations deviating symmetrically by approximately $\pm 5^\circ$ from $\left[1\bar{1}0\right]$. For Pd and Cu, an additional diffraction ring indicates a wide variety of other orientations. An intriguing feature of h-BN on Pt(110) reported by Steiner et al. \cite{Steiner19} is that Pt(110) diplays a $(1 \times n)$ missing row reconstruction with $n$ varying between $5$ and $6$ instead of a $( 1 \times 2)$ missing row reconstruction observed for the clean system \cite{Robinson83}. The modified reconstruction is in registry with the h-BN/Pt(110) moir\'{e} pattern, indicating that the Pt(110) structure responds to the presence of the h-BN overlayer. 

An even more drastic effect of overlayer growth on the substrate structure was found recently by some of us in consequence of low pressure CVD graphene (Gr) growth on Ir(110) \cite{Kraus22}. Under the Gr cover, Ir(110) remains unreconstructed. Instead, without the Gr cover Ir(110) forms upon cool-down below 800\,K $(331)$ nano-facets, with the corrugation of the resulting ridge pattern in the nm range \cite{Koch91}. Similar effects might be caused by the CVD growth of other overlayers, and exploring these effects is promising, particularly given the single crystal quality of graphene on Ir(110). 

In this manuscript, we investigate the growth and structure of h-BN on Ir(110), with focus on analyzing the temperature dependence of orientation order of the h-BN growth, the precise analysis of the emerging moir\'{e}s and the underlayer structure. We find that h-BN growth at a sufficiently high temperature enables a unique epitaxial relation between overlayer and substrate. Moreover, the nano-facet reconstruction known from clean Ir(110) is suppressed, and a $(1 \times n)$ missing row reconstruction, similar to Pt(110), is observed beneath the h-BN layer, though with a different average $n$. Our results are put into context to previous works on other h-BN overlayers on fcc(110) surfaces, and we identify several points of interest for future research by comparing these different systems. 

\section{Results}

\begin{figure}[]
	\includegraphics[width=1\columnwidth]{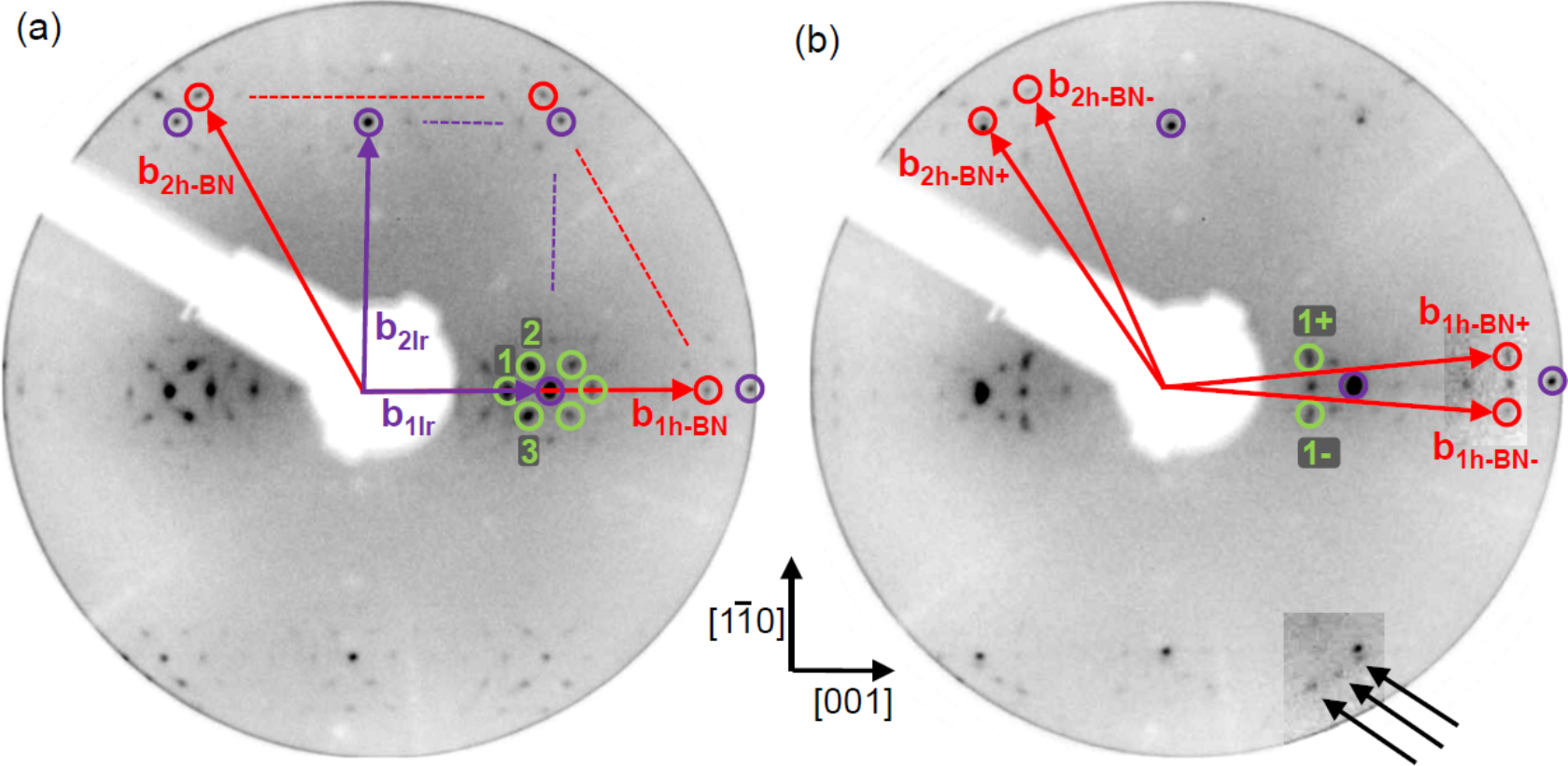}
	\caption{110\,eV LEED patterns after h-BN growth on Ir(110) at (a) 1500\,K and (b) 1250\,K. In (a) some Ir(110) substrate reflections are surrounded by purple circles and the primitive substrate reciprocal translations $\vec{b}_{\rm{1Ir}}$ and $\vec{b}_{\rm{2Ir}}$ are indicated by purple arrows. Some first order h-BN reflections are surrounded by red circles and the primitive h-BN reciprocal translations $\vec{b}_{\rm{1h-BN}}$ and $\vec{b}_{\rm{2h-BN}}$ are indicated by red arrows. The vectors $\vec{b}_{\rm{1Ir}}$ and $\vec{b}_{\rm{1h-BN}}$ are aligned. A set of six green circles highlights moir\'{e} reflections around the first order substrate spot defining $\vec{b}_{\rm{1Ir}}$. See text. In (b) after growth at 1250\,K additionally h-BN domains tilted in average by $\pm 4.7^{\circ}$ with primitive reciprocal vectors $\vec{b}_{\rm{1h-BN+}}$, $\vec{b}_{\rm{1h-BN-}}$, $\vec{b}_{\rm{2h-BN+}}$, and $\vec{b}_{\rm{2h-BN-}}$ are present. The three black arrows in the contrast enhanced part of the LEED pattern highlight the presence of the three domains differing in orientation. New moir\'{e} reflections linked to the tilted domains are encircled green and labeled as $1+$ and $1-$. See text.}
\label{fig1}
\end{figure}

To investigate the structure of h-BN layers grown on Ir(110), it is suitable to start with the analysis of the LEED pattern after growth at 1500\,K as shown in figure~\ref{fig1}(a). It displays the rectangular reciprocal lattice of the unreconstructed Ir(110) substrate, of which the primitive translations $\vec{b}_{\rm{1Ir}}$ and $\vec{b}_{\rm{2Ir}}$ are indicated in purple. Some of the substrate reflections are encircled in purple as well. 

The h-BN lattice can be recognized through its six first order reflections forming a hexagon. Three of these reflections are encircled red and the primitive reciprocal translations $\vec{b}_{\rm{1h-BN}}$ and $\vec{b}_{\rm{2h-BN}}$ are also colored red. Noteworthy, the h-BN lattice is well aligned to the substrate lattice with $\vec{b}_{\rm{1h-BN}}$ being parallel to $\vec{b}_{\rm{1Ir}}$. This alignment implies in real space that the densely-packed rows of Ir(110) are parallel to densely-packed zig-zag directions of the h-BN lattice. The intensity of all six first order h-BN reflections changes simultaneously with energy (data not shown), suggesting that the threefold symmetric h-BN is present in twin domains with the positions of B and N swapped, such that all first order LEED intensities change in the same way with energy. For better distinction from twisted domains, the twinning will be neglected in the following such that twin domains defined by a unique orientation of the zigzag direction will be treated as a single domain.      

All other spots visible in the LEED pattern are moir\'{e} reflections. Six of them are highlighted by green circles, which surround the Ir(110) spot, to which $\vec{b}_{\rm{1Ir}}$ points. The encircled moir\'{e} reflections can be understood to result from fixing the vector differences between first order h-BN reflections and nearby Ir substrate reflections to the first order Ir spot in their center. They can also be constructed from the reciprocal primitive translations. For instance, the moir\'{e} reflections labeled 1, 2, and 3 result from $\vec{b}_{\rm{1h-BN}} -\vec{b}_{\rm{1Ir}}$, $\vec{b}_{\rm{1h-BN}} + \vec{b}_{\rm{2h-BN}} - \vec{b}_{\rm{1Ir}}$, and $\vec{b}_{\rm{2Ir}} - \vec{b}_{\rm{2h-BN}}$. All other reflections not encircled are higher order moir\'{e} reflections around the same Ir(110) spot or moir\'{e} reflections surrounding other Ir substrate spots. The moir\'{e} reflections on the left-hand side of the LEED pattern, not encircled but equivalent to the encircled ones, are slightly streaky indicating some orientation scatter of the moir\'{e}.

If the growth temperature is lowered to 1250\,K, besides aligned h-BN also h-BN twisted by $\pm (4.7 \pm 1.0)^\circ$ in both angular directions becomes visible as obvious from the LEED pattern of figure~\ref{fig1}(b). Instead of one, now three nearby first order h-BN reflections located on a circle segment around the origin are visible. The three black arrows highlight such a first order reflection triplet in a contrast enhanced part of the LEED pattern. Note that one spot in that triplet is very close, but distinct from an Ir substrate reflection. This implies that h-BN zigzag rows for the two domains almost coincide with the $\{1 \bar{1} 2\}$ directions of the substrate. The spots are slightly extended along the circle on which they are positioned, giving rise to the angular scatter specified above. However, the three spots on the segment are clearly distinct. The first order reflections related to the new primitive translations $\vec{b}_{\rm{1h-BN+}}$, $\vec{b}_{\rm{2h-BN+}}$, $\vec{b}_{\rm{1h-BN-}}$, and $\vec{b}_{\rm{2h-BN-}}$ are encircled red. While the moir\'{e} and specifically higher moir\'{e} reflections of the aligned h-BN are now less prominent, new moiré reflections related to twisted h-BN appear. Two of these moir\'{e} reflections are encircled green and are labeled $1+$ and $1-$. They can be expressed as $\vec{b}_{\rm{1h-BN+}} -\vec{b}_{\rm{1Ir}}$ and $\vec{b}_{\rm{1h-BN-}} -\vec{b}_{\rm{1Ir}}$, respectively.      

In conclusion at 1500\,K growth temperature aligned h-BN forms that displays zig-zag directions of h-BN parallel to the Ir(110) dense-packed rows along $\left[1 \bar{1} 0 \right]$. Upon lowering the growth temperature to 1250\,K, aligned and $\pm\,4.7^\circ$ twisted h-BN coexist.For aligned and twisted h-BN the presence of moir\'{e} reflections in LEED indicate the existence of superstructures resulting from the interaction of the h-BN layer with the substrate. For growth temperatures in the range from 1450\,K to 1050\,K the LEED pattern remains similar to the 1250\,K LEED pattern, though it gets rather fuzzy at 1050\,K (not shown). At a growth temperature of 850\,K no sign of h-BN layer formation can be detected in LEED.

\begin{figure}[]
	\includegraphics[width=1\columnwidth]{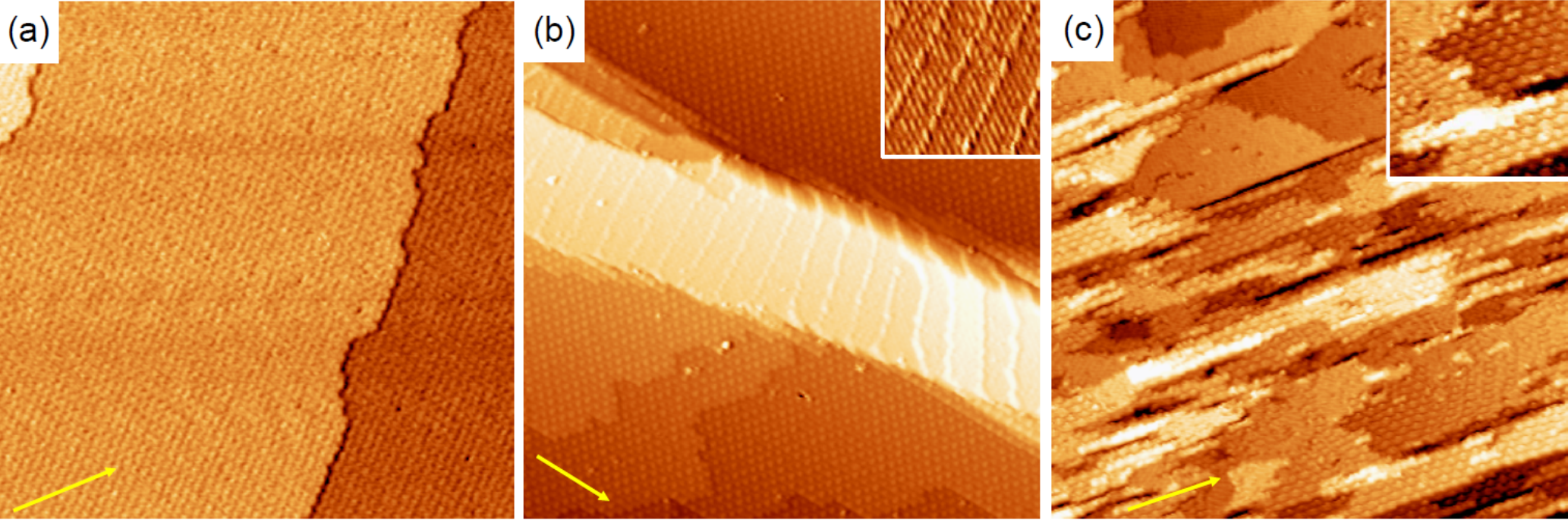}
	\caption{110\,nm $\times$ 110\,nm STM topographs after h-BN growth at (a) 1500\,K, (b) 1250\,K and (c) 1050\,K. Constrast enhanced insets in (b) and (c) point out additional line corrugation and patches of quasi-hexagonal pattern, respectively. Insets in (b)  and (c) have sizes of 27\,nm $\times$ 27\,nm and 30\,nm $\times$ 30\,nm, respectively. (a) and (c) are taken at 300\,K, (b) at 4.2\,K. The yellow arrow in each topograph indicates the $\left[1\bar{1}0\right]$ direction. See text.}
\label{fig2}
\end{figure}

A complementary real-space view of h-BN layer formation as a function of temperature is provided by the sequence of STM topographs in figure~\ref{fig2}.
As visible in figure~\ref{fig2}(a), after growth at 1500\,K the sample is uniformly covered with a quasi-hexagonal pattern of periodicity $\approx 3$\,nm. The sample is rather flat with a low density of mostly monatomic steps. Upon lowering the growth temperature to 1250\,K, the morphology is more rough and displays two distinct surface patterns in figure~\ref{fig2}(b). In the upper and lower part of the topograph the same quasi-hexagonal pattern is visible that was present already after growth at 1500\,K. A bright ridge from upper left to lower right displays a different structure. Bright stripes spaced about $\approx 7$\,nm and almost normal to the $\left[1\bar{1}0\right]$ direction cover the ridge separating the two terraces. As highlighted by the inset, line contrast inclined with respect to the prominent stripes with a periodicity of $\approx 1.3$\,nm is present as well. Compared to the LEED patterns taken after growth at 1500\,K and 1250\,K, one has to conclude that the quasi-hexagonal pattern is related to aligned h-BN with its quasi-hexagonal moir\'{e}, while the stripes and lines on the ridge after growth at 1250\,K must be attributed to twisted h-BN. This conclusion will be substantiated in the subsequent analysis below. Figure~\ref{fig2}(c) displays the morphology after h-BN growth at 1050\,K. The sample is rough on smaller scale, with small terraces separated by steps predominantly oriented along the $\left[1\bar{1}0\right]$ direction. The terraces display either the quasi-hexagonal pattern or appear relatively smooth, but often line contrast is visible, similar to figure~\ref{fig2}(b). The small terrace size is consistent with the blurred LEED pattern at 1050\,K.

\begin{figure}[]
	\includegraphics[width=1\columnwidth]{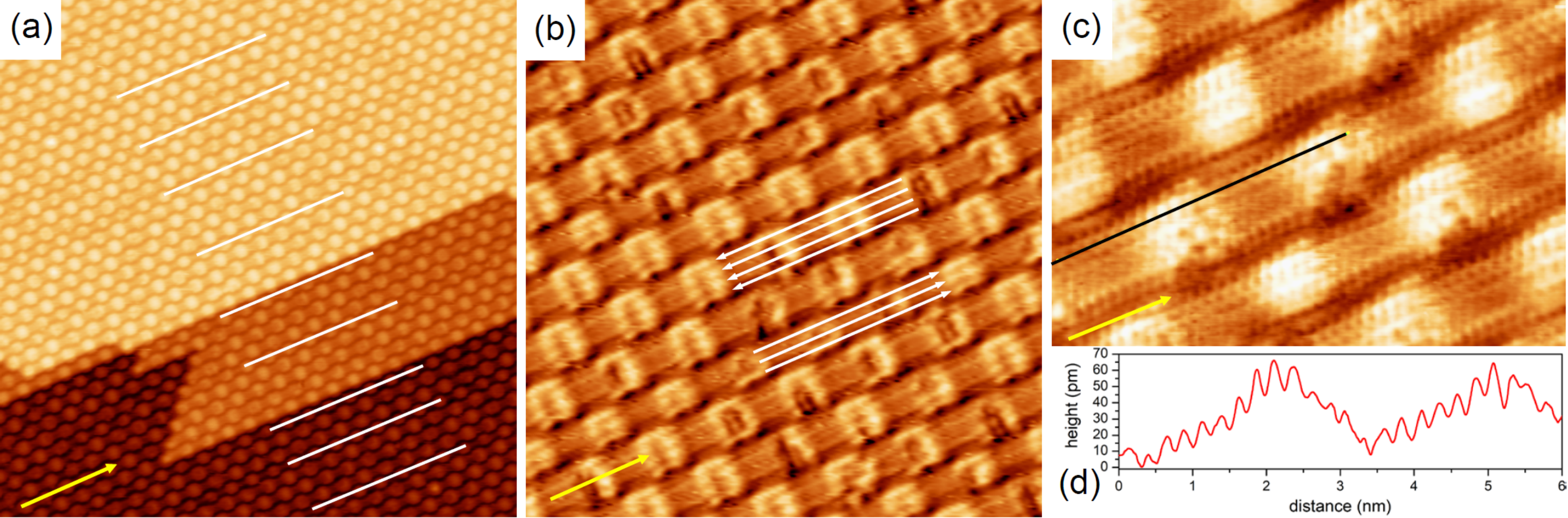}
	\caption{STM topographs of aligned h-BN on Ir(110) after growth at 1250\,K. (a) Topograph displays pronounced dot chains along $\left[1\bar{1}0\right]$ direction. Chains of smaller width are highlighted by white lines. Chains appear to be separated by trenches. (b) White arrows highlight that either four or three bright lines are present in the dot chains. (c) Atomic resolution STM topograph of the h-BN lattice on Ir(110) obtained at very low tunneling resistance ($I_{\rm{t}} = 5$\,nA, $U_{\rm{t}} = -0.25$\,V). (d) Height profile along black line in (c), parallel to the $\left[1\bar{1}0\right]$ direction. The yellow arrow in each topograph indicates the $\left[1\bar{1}0\right]$ direction.  Image sizes are (a) 65\,nm $\times$ 65\,nm, (b) 25\,nm $\times$ 25\,nm, and 9\,nm $\times$ 6\,nm. All topographs taken at 300\,K. See text.}
\label{fig3}
\end{figure}

In the following we examine first the aligned h-BN with its quasi-hexagonal pattern in more detail. The higher resolved topograph of figure~\ref{fig3}(a) makes obvious that the quasi-hexagonal pattern displays a pronounced chain structure of bright dots along the $\left[1\bar{1}0\right]$ direction. The chains appear to be separated by trenches. Two types of chains are present: wider chains with a width of $\approx 1.9$\,nm and narrow chains with a width of $\approx 1.5$\,nm. The later are highlighted in figure~\ref{fig3}(a) through white lines. From the uneven spacing of the white lines it is clear that the distribution of the wide and narrow chains is non-uniform: two narrow chains are separated by two, three, or four wide chains. Also apparent from figure~\ref{fig3}(a) is the effect of the quasi-hexagonal pattern (or the moir\'{e}) on the substrate. Steps are defined precisely through the pattern of bright dots. At upper and at lower edges always a complete chain of bright dots is found. It is thus evident that the substrate reorganizes during growth to maximize bonding between the h-BN layer and Ir(110). 

Figure~\ref{fig3}(b) displays the internal structure of the chains and its relation to the substrate in more detail. Under the tunneling conditions chosen the h-BN layer becomes semi-transparent and tunneling is modulated by the metallic substrate. The white arrows highlight that the bright dots in the wide chains display four rows, the bright dots in the narrow chains exhibit only three rows along the $\left[1\bar{1}0\right]$ direction. These rows are spaced by about $0.4$\,nm and can be identified with the densely-packed Ir(110) rows which are spaced by $0.384$\,nm. The presence of these rows between the bright dots is less clear from the topograph. Nevertheless, the dark line between two chains makes evident that a complete row of Ir is missing. 

When the tunneling resistance is substantially reduced, thereby bringing the tip in close proximity if not contact with the h-BN layer, the layer 
itself is atomically resolved. Consistent with our LEED results, densely-packed rows of the h-BN layer are aligned with good accuracy to the Ir(110) densely-packed rows, i.e., the $\left[1\bar{1}0\right]$ direction. The superperiodicity along $\left[1\bar{1}0\right]$ represented by the bright dots can thus be interpreted as a moir\'{e} effect. The line profile along a densely-packed h-BN layer row in figure~\ref{fig3}(d) makes plain that the distance between two moir\'{e} maxima is 12 distances $a_{\mathrm{h-BN}}$. A systematic analysis with even longer atomically resolved chains yields for the moir\'{e} periodicity $a_{\rm{m}} = (12.0 \pm 0.2) a_{\rm{h-BN}}$. Using the one-dimensional moir\'{e} construction 
\cite{ndiaye08} we obtain 

\begin{eqnarray}
\frac{1}{a_{\mathrm{m}}} = \frac{1}{n\,a_{\mathrm{h-BN}}} = \frac{1}{a_{\mathrm{h-BN}}} - \frac{1}{a_{\mathrm{Ir}}}
\end{eqnarray}

and thus 

\begin{eqnarray}
a_{\rm{h-BN}} = a_{\mathrm{Ir}} \frac{n-1}{n}
\end{eqnarray}

With $n = 12.0 \pm 0.3$ and $a_{\mathrm{Ir}} = 0.271472$\,nm \cite{arblaster10} we obtain $a_{\rm{h-BN}} = (0.2489 \pm 0.0006)$\,nm and $a_{\rm{m}} = 2.99 \pm 0.08$\,nm.  Although this precise lattice parameter determination is only for the $\left[1\bar{1}0\right]$ direction, based on the LEED patterns of figure~\ref{fig1} with the h-BN layer first order reflections forming an equal-sided hexagon, it must be concluded to be uniform for all three densely-packed directions of the h-BN layer with good accuracy. 

The results agree within the limits of error with those obtained for h-BN on Ir(111), where 
$a_{\rm{h-BN}} = (0.2483 \pm 0.0006)$\,nm, $n = 11.7 \pm 0.3$, and $a_{\rm{m}} = 2.91 \pm 0.08$\,nm \cite{Farwick16}. Compared to the literature value $a_{\rm{h-BN}} = 0.2505$ for h-BN bulk \cite{paszkowicz02}, the layer is compressed by about 0.6\,\%. The compressive strain likely builds up during the cool-down from the growth temperature and is caused by the difference between the positive thermal expansion coefficient of Ir \cite{arblaster10} and the slightly negative one for h-BN \cite{paszkowicz02,yates75}, similar as for graphene and h-BN on Ir(111) \cite{,ndiaye08,Farwick16}. We point out that the h-BN lattice parameter specified is valid for growth at 1250\,K. The compression present will be affected by the growth temperature, causing slight changes in a$_{\rm{h-BN}}$.  

\begin{figure}[]
	\includegraphics[width=0.65\columnwidth]{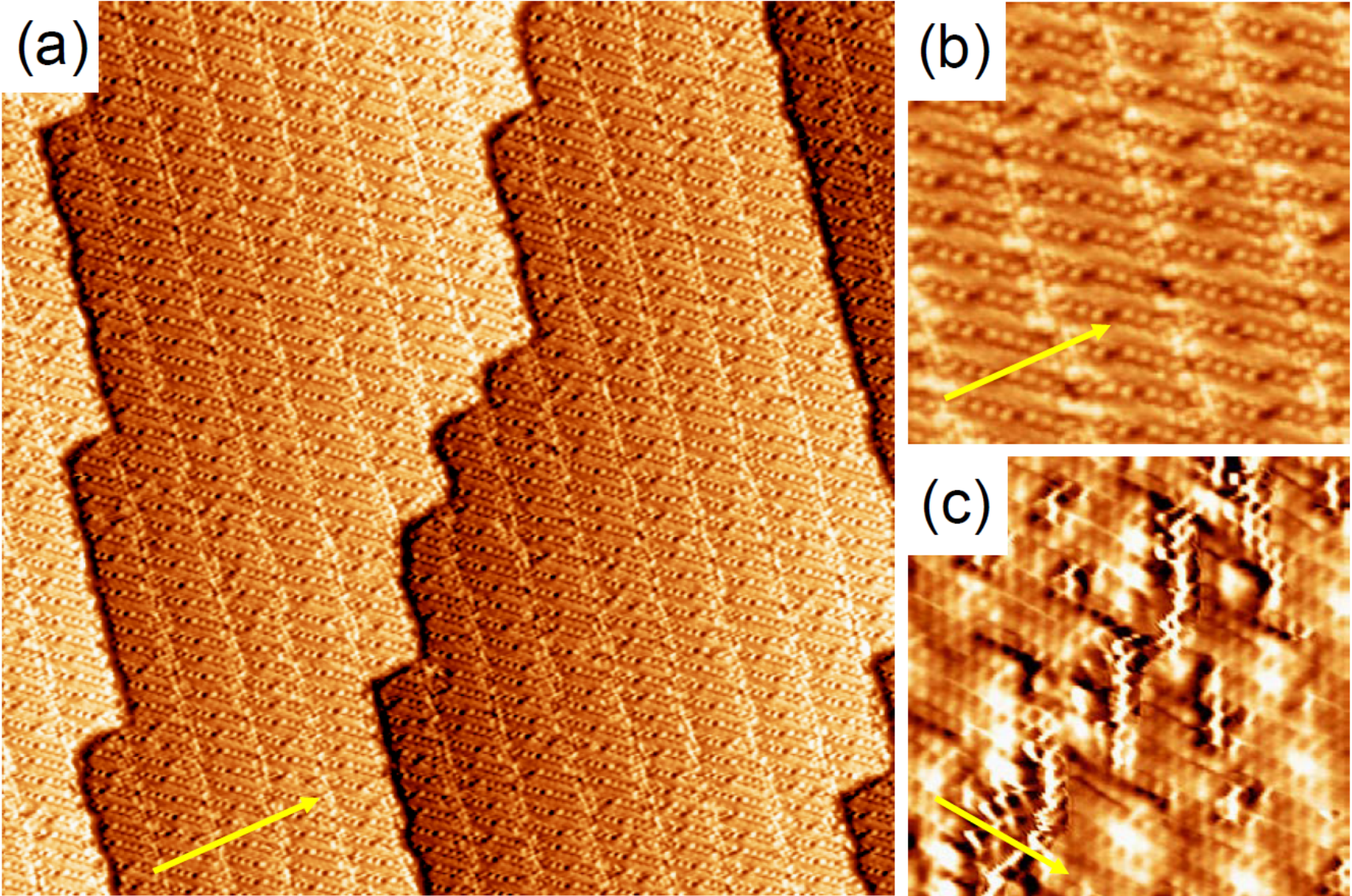}
	\caption{STM topographs of twisted h-BN on Ir(110) after growth at (a),(b) 1450\,K and (c) 1250\,K. (a) displays terraces separated by monatomic steps adjusted to the stripe pattern of twisted h-BN. (b) When zooming into (a) the denser line pattern crossing the stripes becomes visible. (c) Atomic resolution topograph displaying the h-BN lattice of twisted h-BN at low tunneling resistance ($I_{\rm{t}} = 5$\,nA, $U_{\rm{t}} = -0.25$\,V). The yellow arrow in each topograph indicates the $\left[1\bar{1}0\right]$ direction. Image sizes are (a) 75\,nm $\times$ 75\,nm, (b) 20\,nm $\times$ 20\,nm, and (c) 10\,nm $\times$ 10\,nm. (a) and (b) are taken at 300\,K, (c) is taken at 4.2\,K. See text.}
\label{fig4}
\end{figure}

In the following, we turn our attention to twisted h-BN on Ir(110). Figure~\ref{fig4}(a) displays terraces with a twisted h-BN layer extending over the entire topograph. The stripes with a spacing of 6.5\,nm are almost normal to the $\left[1\bar{1}0\right]$ direction and pattern also the step edges. It is obvious that the interaction of a twisted h-BN layer and the Ir(110) substrate cause a reorganization of the Ir(110) substrate such that the moir\'{e} stripes coincide with substrate steps. Also on the ridge covered by twisted h-BN in figure~\ref{fig2}(b), right handside, monatomic steps coincide with stripes. This reorganization must take place during growth and/or at elevated temperatures during the cool-down where Ir diffusion is still active. 

Lines inclined at large angle with respect to the stripes are present in figure~\ref{fig4}(a), but are much better visible in the zoomed topograph of figure~\ref{fig4}(b). The lines are spaced in this case by 1.4\,nm. Additional small bright dots are visible in between, presumably related to the Ir(110) structure underneath. In the atomically resolved and higher magnified STM topograph of figure~\ref{fig4}(c) the atomic structure of the h-BN layer is resolved. Although the image displays substantial additional corrugation at the location of the stripe passing from top right to bottom left, based on the clear LEED pattern related to the twisted h-BN layer in figure~\ref{fig1}(b), we conclude that the h-BN lattice is continuous. Its continuity is difficult to recognize figure~\ref{fig4}(c) due to the additional stripe corrugation related to the interaction of the layer with Ir(110). 

\begin{figure}[]
	\includegraphics[width=1\columnwidth]{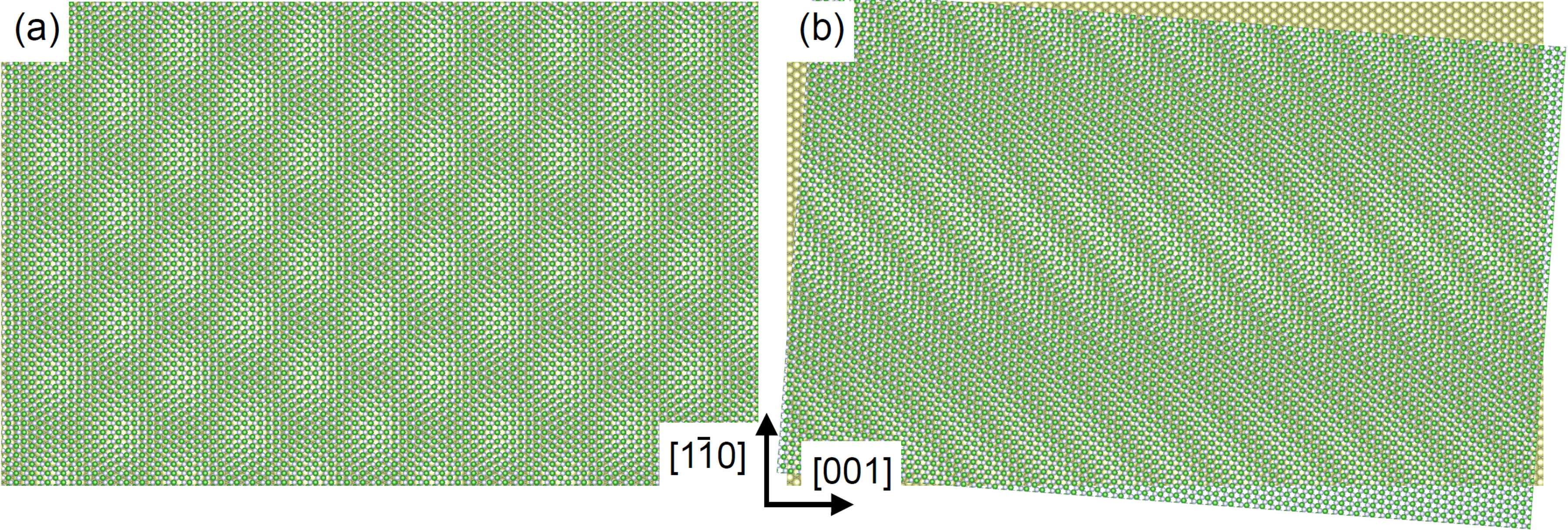}
	\caption{Moir\'{e} resulting from the superposition of a monolayer h-BN with Ir(110). The experimentally determined value $a_{\rm{h-BN}} = 0.2489$\,nm is used as h-BN lattice parameter. In (a) the two lattices are aligned, in (b) they are twisted by 4.5$^\circ$ with respect to each other. See text.}
\label{fig5}
\end{figure}

Figure~\ref{fig5} presents ball models overlaying an h-BN layer with the experimentally determined h-BN lattice parameter on an unreconstructed Ir(110) lattice. For aligned h-BN the quasi-hexagonal 
moir\'{e} pattern is accurately reproduced in figure~\ref{fig5}(a). For twisted h-BN 
figure~\ref{fig5}(b) represents as an exemplary case the moir\'{e} resulting from a twist of 4.5$^\circ$. For this angle horizontal dark stripes almost normal to the $\left[1\bar{1}0\right]$ direction are present with a periodicity of about 6.5\,nm, as well as bright lines spaced at large angle with a distance of about 1.3\,nm. Comparison with figure~\ref{fig2}(b) and figure~\ref{fig4}(c) makes plain, that the experimentally observed periodicities of stripes and lines in twisted h-BN are well reproduced by the ball model. Note that a ball model can only mimic orientations and periodicities of the experiment, but not the STM contrast. Variation of the twist angle in the ball model by just 1$^\circ$ causes large variation of stripe and line orientations as well as changes in the stripe periodicity in the range from 5\,nm to 10\,nm. This explains explain the experimentally observed variation in these quantities. From the good agreement of the ball model moir\'{e}s with our experimental observations we conclude that a largely intact h-BN layer interacts with an Ir(110) substrate still reflecting essential properties of unreconstructed Ir, despite our observation of Ir(110) substrate reorganization. 

Considering the elements of disorder and the limited amount of information extractable from STM topographs about the structure of Ir(110) underneath twisted h-BN, it seems currently not feasible to sketch a model with the details of the substrate structure. The situation is better for the aligned h-BN, which  is characterized by chains of moir\'{e} maxima separated by missing Ir rows. From the analysis of h-BN on Ir(111) \cite{Farwick16} it is known that the interaction of h-BN with Ir is strongest when N-atoms are located on Ir atop sites. This coordination is only possible locally, presumably defining the strongly bound areas in the moir\'{e}. Ir atoms can then be removed from weakly bound areas without loss of much binding. This might explain the locations of removed rows, where binding between h-BN and Ir(110) is weak, and would imply that the bright moir\'{e} maxima are strongly bound, although it does not explain why rows are removed.
The apparent irregularity in the row removal pattern noticed in the discussion of figure~\ref{fig3} would then be simply dictated through the incommensurability of the h-BN lattice with the Ir row periodicity, causing the rows of minimum binding to h-BN being not located at a small integer distance of rows. Slightest deformations of the h-BN lattice due to the details of the step structure as well as strain build-up in the presence of diffusion during the initial stages of cool may also contribute to these irregularities.

\begin{figure}[]
	\includegraphics[width=0.5\columnwidth]{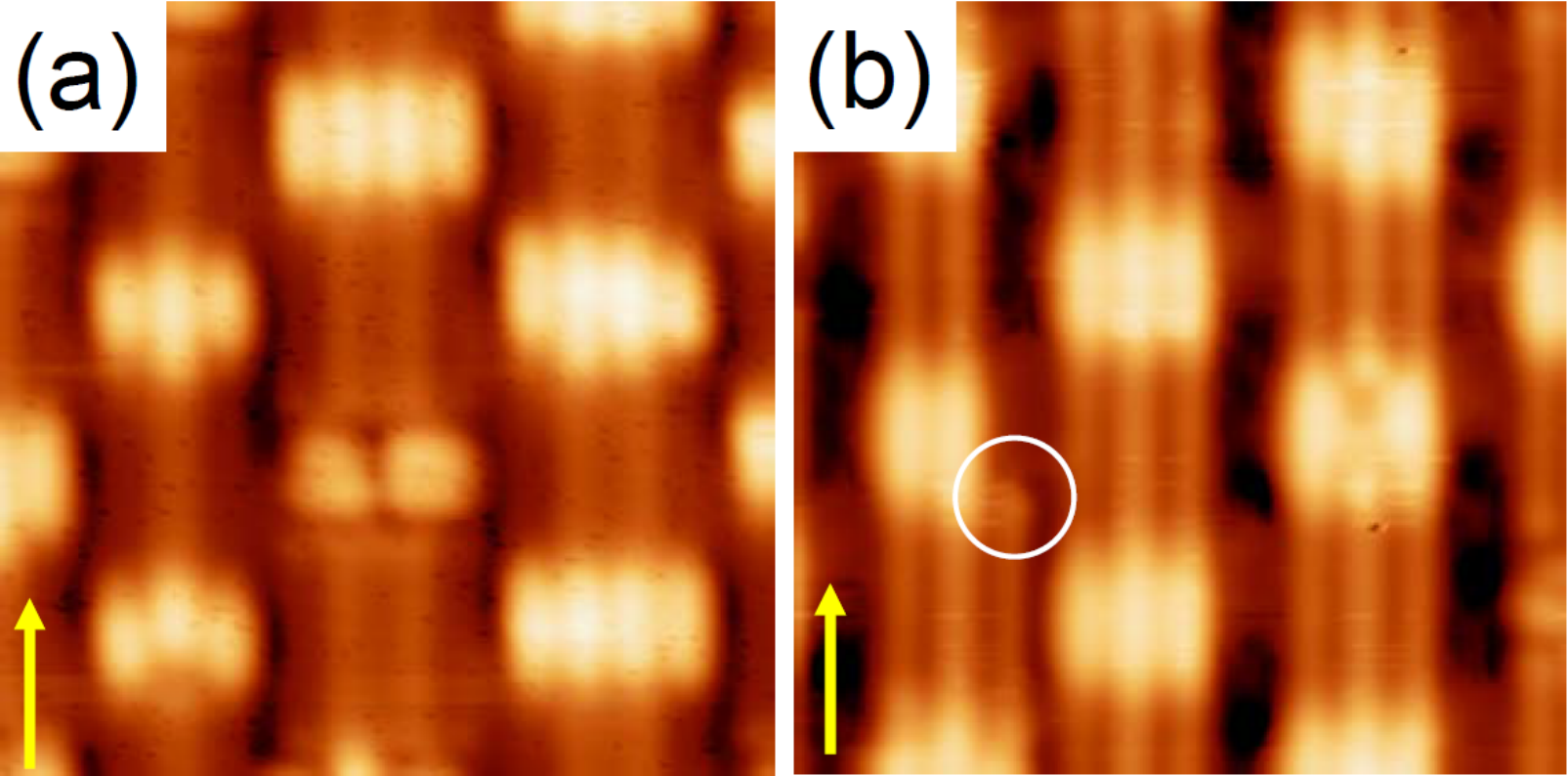}
	\caption{STM topographs of aligned h-BN on Ir(110) after growth at 1250\,K taken at 4.2\,K. While in (a) a single atomic Ir rows appear to be missing between two chains of bright moir\'{e} spots, in (b) mostly two atomic rows are missing. A location where an Ir row terminates is highlighted in (b) through a white circle.  Image size is 6\,nm $\times$ 6 for both topographs. See text.}
\label{fig6}
\end{figure}

At this point it needs to be noted that not only single Ir rows are removed between the chains of the bright moir\'{e} as suggested by figure~\ref{fig3}. Figure~\ref{fig6}(a) and (b) display aligned h-BN grown at 1250\,K in one and the same experiment. It is apparent that the periodicity in $\left[001\right]$ direction defined by the chains is the same. However, while in figure~\ref{fig6}(a) single atomic rows appear to be missing between sets of three or four rows, in figure~\ref{fig6}(b) mostly two rows are missing between sets of two and three Ir rows underneath the h-BN layer. It is, therefore, quite evident that although the clear principle of removing Ir rows is consistently followed, the specifics and quantity of rows removed are influenced by experimental or morphological subtleties that are not under full experimental control. 

\begin{figure}[]
	\includegraphics[width=0.5\columnwidth]{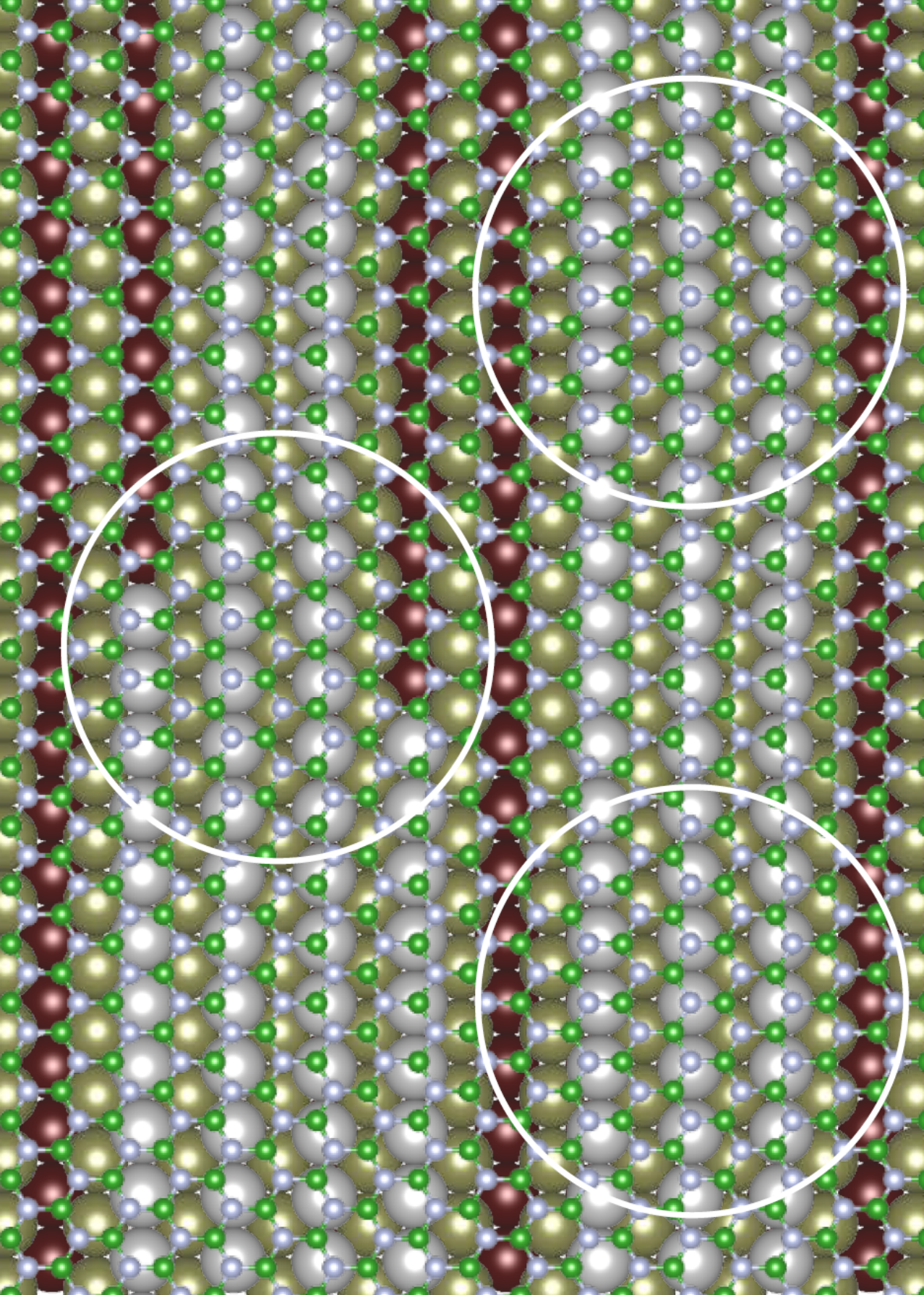}
	\caption{Ball model for aligned h-BN on Ir(110). Dark brown balls: atoms of lowest Ir(110) layer; gold balls: atoms of middle height Ir(110) layer; light gray balls: atoms of topmost Ir(110) layer; small blue balls: N atoms of h-BN layer; small green balls: B atoms of h-BN layer. White circles indicate areas where N atoms sit atop of Ir atoms. Note that the ball model does not provide a unit cell of the surface structure. See text.}
\label{fig7}
\end{figure}

The ball model of figure~\ref{fig7} visualizes our discussion for aligned h-BN. While the upper part sketches the situation when two rows are removed between chains, the lower part sketches the situation with one removed Ir row between two chains. In both cases wide and narrow chains are visible, though their left-right position is swapped. Areas with nitrogen atoms centered atop of Ir atoms must be considered as strongly binding h-BN areas \cite{Farwick16,Steiner19} and are highlighted in figure~\ref{fig7} by white circles. Presumably these areas are the bright dots in the chains when imaged by STM. This speculation is underpinned by the often defective appearance of the bright dots in our STM topographs, since it was shown for h-BN on Ir(111) that h-BN is first damaged by tunneling or temperature treatment in the strongly bonding areas 
\cite{Farwick16}. Note also that for h-BN the strongly bonding areas may appear higher as the weakly bonding ones \cite{schulz14}, making it plausible that also for h-BN on Ir(110) this may be the case. Areas where N atoms are not located atop of Ir(110) are then less strongly bound, enabling the removal of complete rows between the strongly bonding areas. As incomplete removal of rows in between two strongly binding areas would cause energetically costly row terminations, it is well understandable that the rows are continuous along $\left[1\bar{1}0\right]$ between strongly binding areas. 

\begin{figure}[]
	\includegraphics[width=0.45\columnwidth]{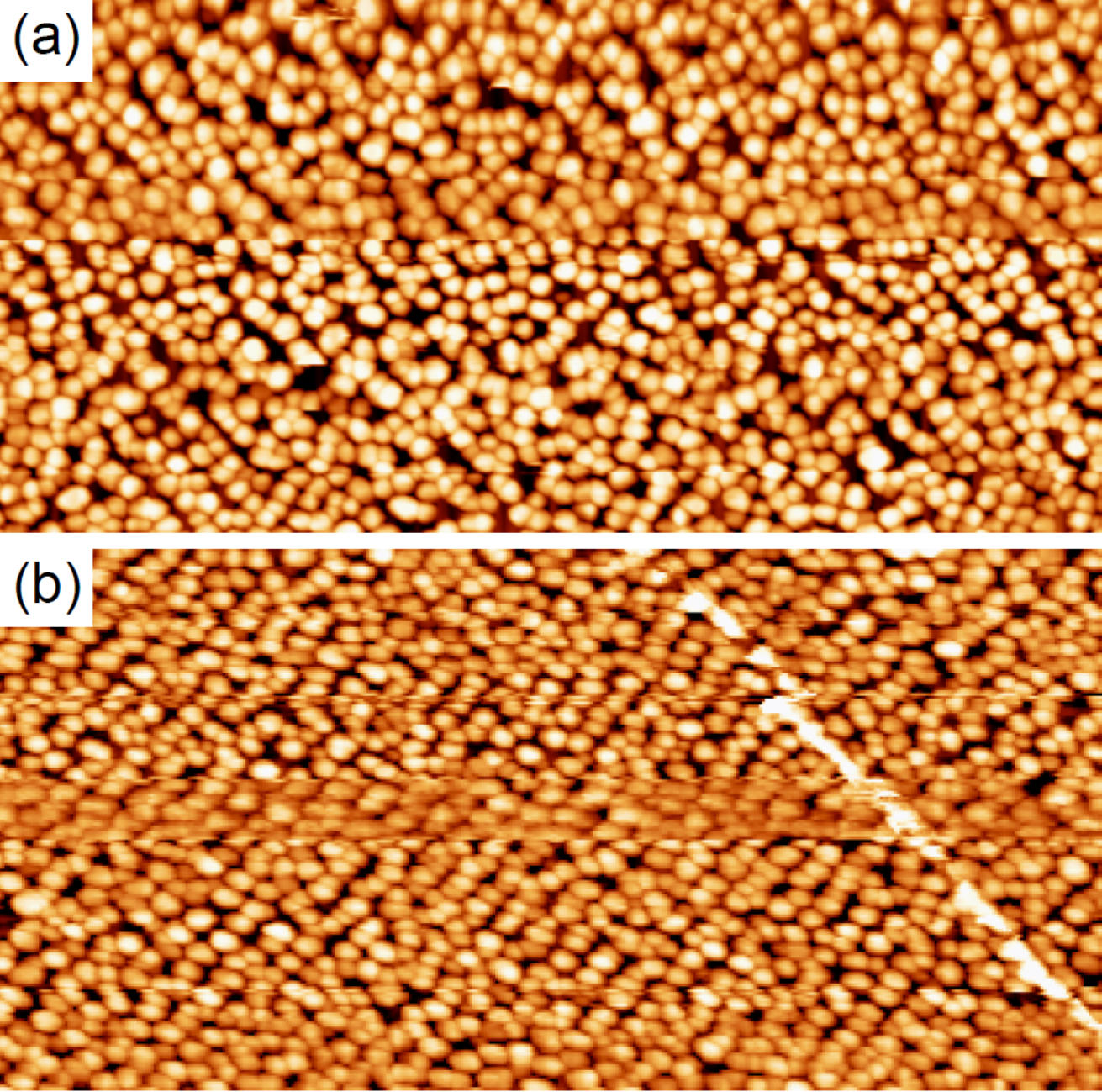}
	\caption{Room temperature STM topographs after deposition of 0.9\,ML Ir on h-BN/Ir(111) at 200\,K. Scan width is 170\,nm.}
\label{fig8}
\end{figure}

Finally, we tested whether the h-BN/Ir(110) moir\'{e}s have the potential to template cluster formation, similar to h-BN/Ir(111) \cite{will18}. As apparent from figure~\ref{fig8}, Ir clusters with typical height in the range 1-1.5\,nm are observed, and they are pinned with a stripe periodicity. In figure~\ref{fig8}(a), the stripe periodicity is 7\,nm, while in figure~\ref{fig8}(b), it is 5\,nm. From the measured periodicity, it is evident that the clusters are pinned on twisted h-BN domains, which display a stripe periodicity in that range together with large periodicity scatter due to the orientation scatter of $\pm 1.0^\circ$ for twisted h-BN/Ir(110) around the average of $\pm 4.7^\circ$. Although the clusters do not form a cluster superlattice as in the case of h-BN/Ir(111), a clear templating effect is manifested through the topographs shown in figure~\ref{fig8}. Exploring the use of twisted h-BN on Ir(110) for templating other species, such as molecules, appears to be a promising avenue for further investigation.

We could not detect templating by aligned h-BN/Ir(110), similar to the results of Kerschbaumer et al. in ref. \citenum{Kerschbaumer22} for Cu deposited on aligned h-BN/Pt(110).

\section{discussion}

The temperature dependence of h-BN growth on Ir(110) closely resembles the temperature dependence of graphene (Gr) growth on Ir(110). At 1500\,K both 2D-materials grow as single domain layers, while at lower temperature they develop additional domains oriented by $\pm 4.7^\circ$ (h-BN) or $\pm 5.4^\circ$ (Gr) with their zigzag directions away from $\left[1\bar{1}0\right]$ (within the limits of error the orientations agree). Both materials suppress the formation of the nano-facet reconstruction upon cool down from the growth temperature, as discussed already in ref. \citenum{Kraus22}. The presence of the 2D-layer makes the undercover surface more bulk-like in terms of bonding and the energy cost for bending and stretching the overlayer due to nano-faceting impedes it \cite{Kraus22}. The general tendency that with increasing growth temperature a 2D layer grown on Ir, be it the (111) or the (110) surface, aligns to a single substrate direction \cite{Hattab11,Kraus22,Farwick16} is obeyed for h-BN on Ir(110) as well. The generality of this observation indicates that with increasing growth temperature kinetics does not hinder or limit the approach of the global maximum in binding energy between substrate and 2D-layer. 

There are also differences: The single domain layer of Gr on Ir(110) makes an angle of $90^\circ$ with respect $\left[1\bar{1}0\right]$, while the single domain layer of h-BN is aligned. The high temperature single crystal graphene is absent upon lowering the growth temperature to 1300 \,K and only twisted Gr is formed. To the contrary, for h-BN twisted h-BN coexists with untwisted h-BN in a broad temperature range from 1050\,K to 1450\,K. The origin of these difference is probably not easy to uncover and beyond the scope of the present manuscript.

The system most similar to h-BN/Ir(110) is h-BN/Pt(110). For both systems single domain h-BN aligned to $\left[1\bar{1}0\right]$ is observed and they are the only systems known to exhibit this behavior (compare Table~\ref{table}). Both systems display a similar $(1 \times n)$ missing row reconstruction under h-BN, with n being in average $\approx 5.5$ for h-BN/Pt(110) and $\approx 4.75 \pm 0.15$ for h-BN/Ir(110). As the rows are removed from underneath the h-BN layer at locations between moir\'{e} maxima, the difference in $n$ reflexts the difference in lattice mismatch between h-BN and the two metals. The variability of the missing row reconstruction seems to be larger on Ir(110) with a less regular spacing of the missing rows and situations where two instead of one row is missing.

Steiner et al. \cite{Steiner19,Steiner23} measure for the $(1 \times 5.5)$ missing row reconstruction by STM height profiles with 9\% to 17\% increased spacing between neighboring Pt rows, depending on the preparation condition. They interpret the measured increase to be physical. For h-BN/Ir(110) we measure also increased spacings between neighboring Ir rows in the order of 10\%. However we consider such increased spacings in STM height profiles as STM imaging artefacts linked to imaging a corrugated substrate through the h-BN cover. The Ir(110) first order reflections in Figure \ref{fig1} are sharp and at the expected crystallographic locations. This would not be the case, if the row spacing would display variability or differences between first and second layer. 

Considering the data provided in the Table~\ref{table}, it is surprising that the moir\'{e} periodicity assumed or indicated for h-BN/Pt(110) varies by about 0.55\,nm, ranging between 2.22\,nm \cite{Achilli18} and 2.78\,nm \cite{Steiner19}. The variation seems to be linked to the different supercells assumed for the density functional theory calculations, rather than to be real differences. A moir\'{e} construction as expressed through equations (1) and (2) should readily resolve these discrepancies.

A last difference between the two systems refers to the temperature dependence of the morphology. While for h-BN/Pt(110) with decreasing temperature the underlayer terrace size decrease and makes the entire system rougher without significant change in domain orientation \cite{Steiner19,Steiner23}, for h-BN/Ir(110) the twisted h-BN domains start to appear already by lowering the growth temperature from 1500\,K to 1450\,K. Significant decrease of underlayer terrace size only becomes noticeable for growth at 1050\,K. This implies that the formation of twisted h-BN on Ir(110) is not linked to a decreasing terrace size.

The formation of twisted h-BN on Ir(110) is similar to what is observed for growth on Pd(110), Rh(110) and 
Cu(110). For all four systems twisted domains form that are with their zigzag direction $\pm 4^\circ$ to $\pm 6^\circ$
off from $\left[1\bar{1}0\right]$ \cite{Corso05, Galera19b, Herrmann18, Herrmann20}. 
In that sense h-BN growth on Ir(110) stands in phenomenology just between 
what is observed for Pt(110) and for the group Pd(110), Rh(110) and Cu(110).

The common characteristics of these twisted h-BN layers is that three of their first order reflections coincide or nearly coincide with a $\left\{ 1 1 \right\}$-type reflection, i.e., in Figure \ref{fig1}b the h-BN spot at  $- \vec{b}_{\rm{2h-BN+}}$ nearly coincides with the Iridium spot at $\vec{b}_{\rm{1Ir}}$ - $\vec{b}_{\rm{2Ir}}$ (the rightmost of the three black arrows on the bottom right highlights this location). The same near coincidence of h-BN and substrate reflections is visible also for Pd(110) \cite{Corso05}, Rh(110) \cite{Galera18, Galera19b} and Cu(110) \cite{Herrmann18, Herrmann20}. Translated to real space, this coincidence implies the alignment of the h-BN zigzag direction with the substrate $\{ 1 \bar{1} 2 \}$ directions, that is with the $\left[ 1 \bar{1} 2 \right]$ or the $\left[ \bar{1} 1 2 \right]$.

While it is not obvious why this coincidence is realized by nature, the impressive results obtained by Wang et al. \cite{Wang19} demonstrating single crystal h-BN layer growth on a surface vicinal to Cu(110) become understandable. The authors selected Cu(110) with step bunches along the $\left[ 1 \bar{1} 2 \right]$ direction. Thereby perfect match of $\left[ 1 \bar{1} 2 \right]$ steps with the h-BN zig-zag direction triggers heterogeneous nucleation of h-BN islands at the step-down edges in precisely one of the two domains preferred by the system. Since all steps have the same step-down direction, only one of the two twin domains can be realized and the resulting layer is indeed single crystal. Based on the fact that the twist domains of Cu, Pd, Rh and Ir all show the same properties, we predict that miscut (110) samples with $\left[ 1 \bar{1} 2 \right]$ steps will enable the same single crystal growth.

As can be inferred from Table~\ref{table}, the LEED pattern of h-BN/Ni(110) present in ref. \citenum{Greber06} falls out of the scheme presented here. This is the more surprising, as $a_{\rm{Ni}} = 0.2492$\,nm matches nearly perfectly with the h-BN bulk lattice parameter $a_{\rm{h-BN}} = 0.2505$\,nm such that one would assume the domain with aligned and pseudomorphic zigzag and $\left[1\bar{1}0\right]$ directions to be energetically much more favorable than all other domains. It might be worth to look once more into h-BN/Ni(110) and its temperature dependent formation to understand better the variety of structures reported.

Given that for h-BN/Pd(110) and h-BN/Rh(110) besides twisted h-BN no coexistent aligned h-BN was found \cite{Corso05,Galera18}, there is little chance that by variation of the growth conditions an aligned domain can be created, and even less chance to make this aligned domain the only one. For Cu(110) under certain preparation conditions in fact aligned h-BN was found to coexist with twisted h-BN (compare Figs. 5.19 and 5.20 of ref. \citenum{Herrmann20}). For this system by systematic variation of the growth conditions a specific breeding of aligned h-BN as single domain might be possible.

\begin{table}[h!]
     \begin{center}
	\fontsize{9pt}{9pt}\selectfont
     \begin{tabular}{ | m{1.2cm} | m{3cm} | m{1.7cm} | m{1.8cm} | m{2.5cm} | m{1.7cm} | m{1.1cm} | m{1cm} | m{1.1cm} |}
     \hline
	metal & alignment & single domain & T$_{\rm{growth}}$ (K) & superstructure along $\left[1\bar{1}0\right]$ & a$_{\rm{m}}$ along $\left[1\bar{1}0\right]$ (nm) & a$_{\rm{h-BN}}$ & a$_{\rm{metal}}$ (nm) & ref. \\ \hline
	Ir(110) & $\parallel \left[1\bar{1}0\right]$ and $\pm4.7^\circ$ off & yes & $1050-1500$ & 12.0 on $11.0 \pm0.3$ & 2.99 & 0.2489 & 0.2715 & this work \\ \hline
	Pt(110) & $\parallel \left[1\bar{1}0\right]$ & yes & $1000$ & 9 on 8; c($8\times10$) & 2.216 & 0.2467 & 0.2775 & \cite{Achilli18} \\ \hline
	Pt(110) & $\parallel \left[1\bar{1}0\right]$ & yes & $950-1170$ &	11 on 10 &	2.775 &	0.2523 & 0.2775 & \cite{Steiner19} \\ \hline
	Pt(110) & $\parallel \left[1\bar{1}0\right]$& yes	& 1170 & n/a & 2.5 & n/a & 0.2775 & \cite{Thaler20} \\ \hline
	Pd(110) & $\parallel \left[1\bar{1}0\right]$ and $\pm6^\circ$ off, \newline + diffraction ring & no & 1000 & 11 on 10 & 2.751 & 2.501 & 0.2751 & \cite{Corso05} \\ \hline
	Rh(110) & $\pm6^\circ$ off $\left[1\bar{1}0\right]$ & no & 1070 & n/a & n/a & n/a & 0.2689  & \cite{Galera18, Galera19b} \\ \hline
	Cu(110) & $\parallel \left[1\bar{1}0\right]$ and $\pm5^\circ$ off, \newline + diffraction ring	& no & $970-1020$ & n/a & n/a & n/a & 0.2556 & \cite{Herrmann18, Herrmann20} \\ \hline
	Cu(110) vicinal & +5$^\circ$ off $\left[1\bar{1}0\right]$, \newline zigzag $\parallel \left[1\bar{1}2\right]$ & yes, single crystal & 1310 & n/a & n/a & n/a & 0.2556 & \cite{Wang19} \\ \hline
	Ni(110) & $\parallel \left[1\bar{1}0\right]$, and others & no & 1000 & commensurate & none & 0.2492 & 0.2492 & \cite{Greber06} \\ \hline
      \end{tabular}
      \caption{Table reviewing some growth and structural properties of h-BN/fcc(110) systems. $T_{\rm{growth}}$ denotes the growth temperature, $a_{\rm{m}}$ is the moir\'{e} periodicity, $a_{\rm{h-BN}}$ the lattice parameter of hexagonal boron nitride on the metal, and for reference also the nearest neighbor metal distance $a_{\rm{metal}}$ is provided. See text.}
      \label{table}
      \end{center}
      \end{table}

\section{Conclusions}

In summary, we found that in terms of oriented h-BN growth on fcc(110) surfaces, Ir(110) as a substrate falls between Pt(110) and a group of metals, including Cu, Pd, and Rh. At a growth temperature of 1500\,K, h-BN/Ir(110) displays single domain h-BN aligned with its zigzag directions parallel to the $\left[ 1 \bar{1} 0 \right]$ direction, similar to Pt(110). Between temperatures of 1050\,K and 1450\,K, additional h-BN twisted by $\pm (4.7 \pm 1.0)^\circ$ is observed, similar to what is found for h-BN on Cu, Pd and Ni.

For aligned h-BN, our moir\'{e} analysis yields a moir\'{e} periodicity of $a_{\rm{m}} = 2.99 \pm 0.08$\,nm along $\left[ 1 \bar{1} 0 \right]$ and a h-BN lattice parameter $a_{\rm{h-BN}} = (0.2489 \pm 0.0006)$\,nm, in good agreement with data for h-BN/Ir(111). The slightly smaller lattice parameter of h-BN on Ir(110) compared to bulk h-BN is attributed to thermal compression of Ir during cool-down from the growth temperature.

For twisted h-BN, the observed moir\'{e} pattern with average periodicities of $\approx 7$\,nm for stripes and $\approx 1.3$\,nm for lines displays significant scatter due to the sensitivity to the precise twist angle. The moir\'{e} pattern is well reproduced in a ball model where h-BN is placed on unreconstructed Ir(110). Twisted h-BN imposes the stripe periodicity on the growth of Ir clusters on h-BN/Ir(110), suggesting that also for other species twisted h-BN might be an interesting template.

The h-BN layer on Ir(110) suppresses the nano-facet reconstruction of Ir(110), and under the aligned h-BN a $( 1 \times n)$ missing row reconstruction is formed, with $n = 4.75 \pm 0.15$ and one or two topmost lattice rows removed with this average periodicity. For twisted h-BN, STM does not provide a clear view through h-BN onto the substrate, but the sharp substrate reflections in LEED indicate that the (110) character of the substrate is retained.

Based on the similarities of h-BN domain growth on different fcc(110) surfaces analyzed, we clarify the formation of a single crystal h-BN layer on miscut Cu(110). We predict that the same mechanism of domain twist and twin selection could be operative also for Ir(110), Pd(110), and Rh(110) with miscut exhibiting $\left[ 1 \bar{1} 2 \right]$ steps.  

\section{Methods}

The experiments were conducted in two ultra high vacuum systems (base pressure in the $10^{-11}$\,mbar range) equipped with sample preparation facilities, scanning tunneling microscopy (STM) -- operating either at 300\,K or 4.2\,K -- and low energy electron diffraction (LEED).

Ir(110) was cleaned by cycles of noble gas sputtering (4.5\,keV Xe$^+$ at 75$^\circ$ with respect to the surface normal or with 1\,keV Ar$^+$ at normal incidence) and annealing to 1500\,K. Starting at 1500\,K during cool-down from the last annealing cycle, the sample is exposed to an oxygen partial pressure of $1 \times 10^{-7}$\,mbar to prevent the formation of the $\left\{331\right\}$-nanofacet reconstruction \cite{Koch91}. For growth of the h-BN layer the sample is exposed for 100\,s to a pressure of $1 \times 10^{-7}$~mbar borazine (B$_3$N$_3$H$_6$) at a temperature in the range of 1050\,K to 1500\,K. While the pressure indicated is measured with a distant ion gauge, borazine was admitted through a 10\,mm diameter gas dosing tube ending about 3\,cm in front of the sample with a pressure enhancement factor of more than 20. To test templating of h-BN/Ir(110), Ir was deposited at 200\,K on h-BN/Ir(110) with a flux of $1.4 \times 10^{17}$\,atoms/m$^2$s for 60\,s resulting in a deposited amount of $8.6 \times 10^{18}$\,atoms/m$^2$ or 0.9\.ML, where 1 monolayer (ML) is the surface atomic density of Ir(110).

STM imaging was conducted either at 300\,K or at 4.2\,K and the software WSxM was used for image analysis \cite{horcas07wsxm}. For better visibility of low corrugation features, the STM topoographs in Figs.~2-4 are a superposition of the topograph and its derivative. After the STM measurements, micro-channelplate LEED patterns were digitally recorded over an energy range from about 70\,eV to 200\,eV. The LEED patterns were digitally unwarped to represent the two-dimensional reciprocal lattice without significant distortion.

\begin{acknowledgements}
This work was funded by the Deutsche Forschungsgemeinschaft (DFG, German Research Foundation) -- project number 277146847 -- CRC 1238 (subproject A01). JF acknowledges financial support from the DFG SPP 2137 (Project FI 2624/1-1).
\end{acknowledgements}

\appendix
\bibliography{literatur}
\clearpage

\end{document}